\documentclass[aps,prd,epsf]{revtex4}
\usepackage{graphicx}
\begin{document}
\title{$\rho$ and $\sigma$ Mesons in Unitarized Thermal $\chi$PT}
\author{F.Llanes-Estrada, A. Dobado, A. G\'omez Nicola, J.R.Pel\'aez*}
\address{Deptos. F\'{\i}sica Te\'orica I y II, Univ. Complutense, 28040 Madrid,
Spain.\\  $*$ Also Dip. di Fisica, Universita' degli Studi and INFN,
Firenze, Italy. }
\begin{abstract}
We present our recent  results for
the $\rho$ and $\sigma$ mesons considered as resonances in
pion-pion scattering in a thermal bath. We use  chiral
perturbation theory to order $p^4$ for the low energy behaviour,
then extend the analysis via the unitarization method of the
Inverse Amplitude into the resonance region. The width of the rho
broadens about twice the amount required by phase space
considerations alone, its mass staying practically constant up to
temperatures of order 150 MeV. The sigma meson behaves in
accordance to chiral symmetry restoration expectations.
\end{abstract}
\maketitle

\begin{figure}
\includegraphics[height=.3\textheight]{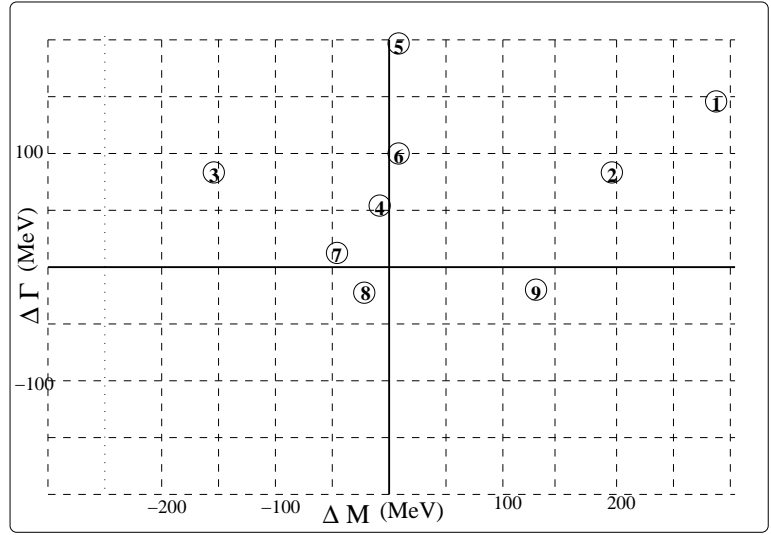}
\caption{
Various existing calculations on the thermal
(at about T=150 MeV) vector meson
mass and width. The numbers refer to the bibliography.}
\label{estanque}
\end{figure}

Triggered by the possibility of exploiting the dilepton spectrum as
a signal of the Quark and Gluon Plasma in Relativistic Heavy Ion
Collisions, there have been numerous studies of the thermal
behaviour of the $\rho$ resonance in a hot hadron medium. Some of
these theoretical approaches are summarized in figure
(\ref{estanque}). The wide spread of predictions signals a strong
model dependence in the various calculations, although a general
(not universal) trend points to a natural broadening of the width
due to a larger available phase space.
In this work we apply chiral perturbation theory \cite{GaLe1984}($\chi$PT)
complemented with the Inverse Amplitude Method \cite{JRPAD} (IAM)
to obtain predictions guided only by the principles of chiral
symmetry and unitarity. 
These combined methods provide outstanding
fits to low energy pion scattering data \cite{JRPAD} thus fixing 
the low energy constants of the chiral lagrangian.
We use this same constants in our thermal amplitudes and all finite temperature
results fall as predictions. 

The starting point is a calculation of the thermal pion scattering
amplitude to one loop in chiral perturbation theory, which
includes calculating pion loops and tadpoles (where temperature
dependence is included through Matsubara sums in the imaginary
time formalism) and polynomial counterterms with arbitrary
coefficients (the low energy constants) which are fitted to known
pion scattering phase shifts. Since the thermal bath induces a
preferred reference frame, our results \cite{ourplb} can be best
written in terms of the $\pi_+ \ \pi_- \longrightarrow \pi_0 \
\pi_0$ scattering amplitude $A({\bf S},{\bf T},{\bf U},\beta)$
which depends on the four vectors ${\bf
S}=p_{\pi_+}+p_{\pi_-}$...  generalizing the Mandelstam
variables and allowing  us to generate
the other possible pion scattering amplitudes by crossing symmetry.

At finite temperature and for $s>4m_\pi^2$ and below other
thresholds, the partial wave projections obtained from our
amplitude satisfy the unitarity relation perturbatively, that is
\begin{equation}\label{unitarity}
{\rm Im}\ a_2(s)=0 \ , \ \ {\rm Im}\ a_4(s;\beta)=\sigma_\beta({\bf S}_0)
\arrowvert a_2(s) \arrowvert^2 \ , \ {\bf S}_0>2m_\pi \ ,
\end{equation}
with a thermal phase space
\begin{equation} \label{phasespace}
\sigma_{\beta}(E)=\sqrt{1-\frac{4m_\pi^2}{E^2}}\left[ 1+
\frac{2}{{\rm exp}(\beta \arrowvert E \arrowvert /2)-1} \right] \ .
\end{equation}

 The temperature is of order $p$, $M_\pi$ in the chiral
counting, and thus for small temperatures and momenta all
corrections scale at least as the fourth power of the 
small quantity $M_\pi/(4\pi
f_\pi)$, providing very good accuracy in
this regime. Alas, the chiral expansion is intrinsically a
polynomial with chiral logarithms (from pion loops)
providing the imaginary parts in (\ref{unitarity}) only perturbatively,
but violating severely exact unitarity already around
the
$\rho$ resonance region.
Still, from the behaviour of the phase shifts at low energy, 
we find a model independent thermal
enhancement  of the low energy interactions, 
conserving their attractive or repulsive nature. This effect is consistent
with
a thermal increase of the rho width
and an almost constant mass\cite{ourplb}. 
To extend this approach to momenta and
temperatures beyond the reach of one loop $\chi$PT, 
several unitarization methods have been proposed.
 Here we employ the well tested IAM (see
\cite{littlebook} for a thorough introduction). In terms of the
$\chi$PT momentum expansion, one approximates $a(s)$ by
\begin{equation}\label{IAM}
a^{\rm IAM}(s;\beta)=\frac{a_2^2(s)}{a_2(s)-a_4(s;\beta)}
\end{equation}
satisfying
\begin{equation}
{\rm Im}\ a^{\rm IAM}=\sigma_\beta(s) \arrowvert
a^{\rm IAM}(s; \beta) \arrowvert^2 \ .
\end{equation}
 One can think of a particular unitarization method giving an
exact treatment of the imaginary part of the inverse amplitude but just an
approximation to the real part; the IAM uses the $O(p^4)$ $\chi$PT
amplitude for the approximation and can be considered as the [1,1]
Pad\'e approximant for the $\chi$PT series in inverse powers of $(4\pi
f_\pi)^2$. Further, the rational formula (\ref{IAM}) allows the
treatment of one pole resonance in each $IJ$ channel which can be
tracked down in the second Riemann sheet for complex $s$. If we
think of the doubly subtracted dispersion relation satisfied by
$a(s)$, the left cut corresponding to $t$-channel pion loops is only
approximately considered, whereas the right $s$-channel cut is
exactly taken into account. This induces a small crossing
violation, but its impact on the resonance region is numerically
controlled because of the large distance of this cut to the $\rho$ or
$\sigma$ poles in the complex plane.

\begin{figure}
\includegraphics[height=.3\textheight]{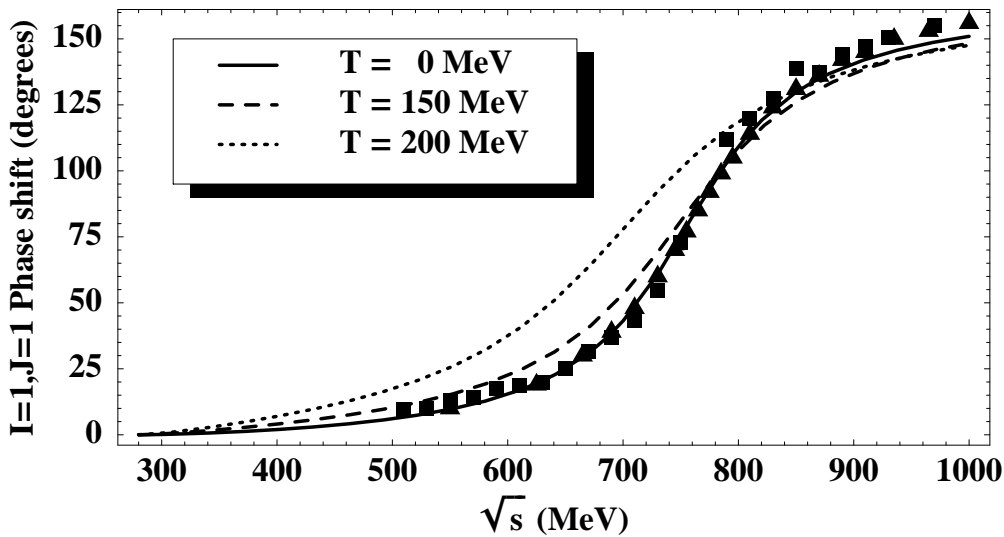}
\caption{Zero and finite temperature pion scattering phase shifts
in the $\rho$ channel.}
\label{phaseshifts}
\end{figure}

\begin{figure}
\includegraphics[height=.3\textheight]{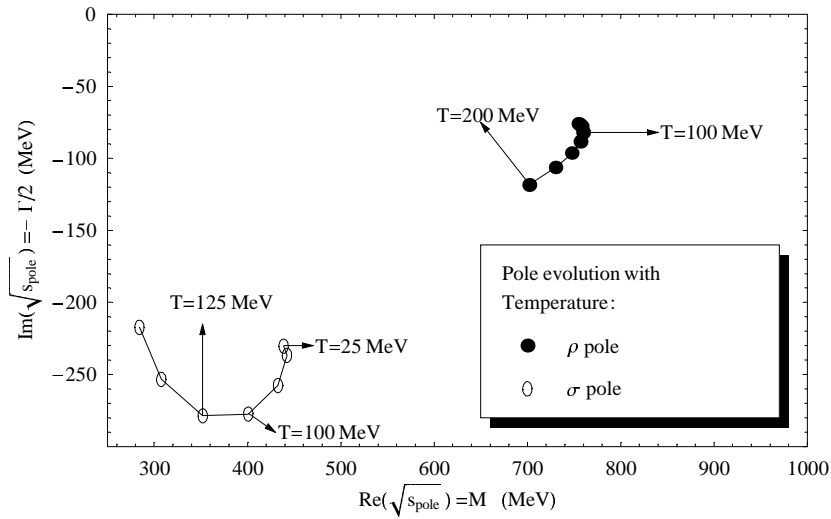}
\caption{How dinamically generated resonance parameters evolve with
the temperature.}
\label{polos}
\end{figure}

Figure (\ref{phaseshifts}) displays the IAM $\rho$ channel phase
shift in pion scattering, at zero temperature (compared to
experimental data) and the finite temperature prediction. The
results are consistent with the evolution of the $\sigma$ and
$\rho$ poles in the complex plane shown in fig.(\ref{polos}) (the
analytical extension of the loop and tadpole integrals to the
complex plane is described in \cite{ourprc}).

The behaviour of the $\sigma$ hints at chiral symmetry
restoration: observe how its mass decreases as the temperature
raises and how its width, initially increasing, dramatically drops
when approaching the relatively stable two pion threshold. On the
other hand, the width of the $\rho$ increases in all the
considered temperature range. This calculation, performed ab
initio in the chirally broken phase, does not have direct access
to the phase transition, but serves to pin down models which do
have this capability providing them with a general behaviour at
low temperatures.

\begin{figure}
\includegraphics[height=.3\textheight]{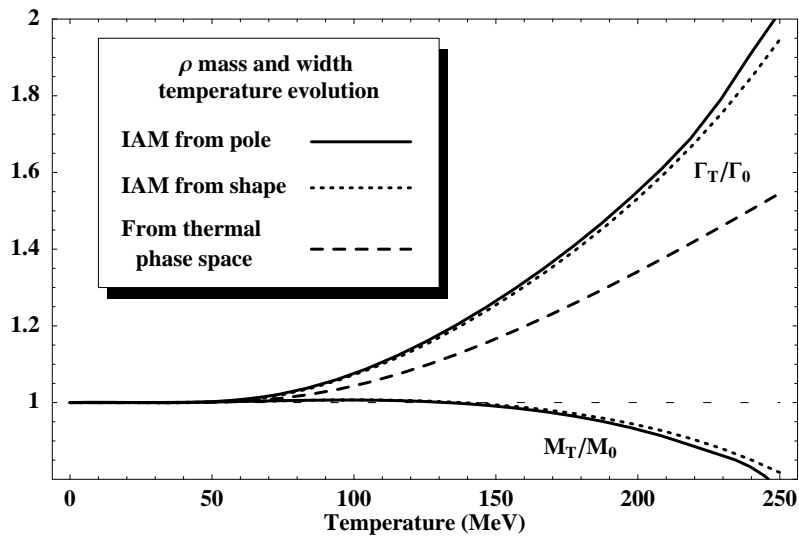}
\caption{Thermal evolution of the $\rho$ mass, width and $\rho \pi \pi$
effective vertex.}
\label{rho}
\end{figure}

Since in our approach resonances are not explicitly included, 
but arise dynamically, we can predict the temperature evolution of
their masses and couplings.
If we 
specifically concentrate on $\rho$ meson properties, replotted
for convenience in figure \ref{rho}, we find that the thermal
width grows faster than the naive expectation from the well known \cite{4}
space phase in eqn.(\ref{phasespace}), the effective $g_{\rho \pi \pi}$
acquires a temperature dependence which was ignored in most of
previous works, but in agreement at low temperatures with an old 
calculation\cite{Song}
and significantly the $\rho$ mass, which stays practically constant,
in agreement with \cite{Ko} and specially \cite{4} (both papers offering
an explanation of the dilepton excess below $M_\rho$). 

{\bf Acknowledgements} Work supported by grants FPA2000-0956,
PB98-0782 and BFM2000-1326. J. R. P. acknowledges also support
from the CICYT-INFN collaboration grant 003P 640.15.


\begin{thebibliography}{20}
\bibitem{3}Same, if VMD does not hold.
\end{thebibliography}
\end{document}